\documentclass[aps,showpacs,twocolumn]{revtex4-2}
\usepackage{epsfig}
\usepackage{graphicx,color,dcolumn}
\usepackage{epstopdf}
\usepackage{amsmath,amssymb}
\usepackage{multirow}
\usepackage{diagbox}
\usepackage{changes}
\usepackage{booktabs}
\usepackage{threeparttable}
\usepackage{subfigure}
\usepackage{hyperref}

\begin{document}
\title{ Systematic study of $D^*_{s0}(2317)$  in quark model}

\author{Yue Tan}
\email[E-mail: ]{tanyue@ycit.edu.cn}
\affiliation{Department of Physics, Yancheng Institute of Technology, Yancheng 224000, People's Republic of China}

\author{Yuheng Wu$^1$}
\email[E-mail: ]{191002007@njnu.edu.cn}
\affiliation{Department of Physics, Yancheng Institute of Technology, Yancheng 224000, People's Republic of China}

\author{Youchang Yang}
\email[E-mail: ]{yangyc@gues.edu.cn (Corresponding author) }
\affiliation{Department of Physics, Guizhou University of Engineering Science, Bijie 551700,  People's Republic of China.}

\author{Jialun Ping}
\email[E-mail: ]{jlping@njnu.edu.cn (Corresponding author)}
\affiliation{Department of Physics, Nanjing Normal University, Nanjing 210023, People's Republic of China}
\date{\today}
\begin{abstract}
%Recently, BES\uppercase\expandafter{\romannumeral3} Collaboration updated the data of exotic state, $D^{*}_{s0}(2317)$, with mass at
%$2318.3\pm1.2\pm1.2$ MeV in the positronium annihilation process. Inspired by experiment, We study the two hot exotic states,
%$D^{*}_{s0}(2317)$ and $D_{s1}(2460)$, in the more reasonable unquenched quark model with two modified factor: the one is  a
%suppression factor, $e^{f^2 {\textbf{p}}^2}$, to reduce the  probability of hadron loops with very high energy,
%while the other factor is a convergence factor, $e^{- \frac{\textbf{r}^2}{4 f^2}}$, which could make the quarks converges to a reasonable range. Different from the other works, we first give all of possible the mass shift and component of
%$D^{*}_{s0}(2317)$ and $D_{s1}(2460)$ at the same time. All the calculation are done with the help of Gaussian expansion method, which
%is very accurate few body system calculation method. In addition, not only $DK(D^*K)$ but also the other OZI allowed mesonium including
%$D_s\eta$ are all taken into consideration. We propose an unquenched quark model to give a unitive description for $D_s$ spectrum including
%four particles: $D_s$, $D_s^{*}$, $D^{*}_{s0}(2317)$ and $D_{s1}(2460)$, and the results show that $D^{*}_{s0}(2317)$ and $D_{s1}(2460)$
%are a $c\bar{s}$ sate  plus an extra important component due to continuum while $D_s$ and $D_s^{*}$ are almost pure quark-antiquark state.
Recently, the BES\uppercase\expandafter{\romannumeral3} Collaboration updated the data for the exotic state, $D^{*}_{s0}(2317)$, with a mass of $2318.3\pm1.2\pm1.2$ MeV from the positronium annihilation process. Inspired by the experiment, we systematically investigated the $c\bar{q}$ meson family spectrum from the perspectives of two-quark, four-quark, and mixed two-four-quark structures, including $D_s$, $D_s^{*}$, $D_{s0}^{*}(2317)$, and $D_{s1}(2460)$, with the help of the Gaussian expansion method, which is a very accurate calculation method for few-body systems. We found that with appropriate spin-orbit coupling parameter $a_s$, the chiral quark model can effectively describe both $D_{s0}^{*}(2317)$ and $D_{s1}(2460)$. The four-quark calculation shows that the $DK$ state does not form a bound state, but the effective potential calculation shows an attractive interaction between them. Finally, we performed mixed two-four-quark calculations, and the results, with the help of the unquenched effect, show that the energies of $D_{s0}^{*}(2317)$ and $D_{s1}(2460)$ are in better agreement with experimental values. Therefore, we conclude that these two particles are good candidates for mixed two-four-quark states.
\end{abstract}

%\pacs{13.75.Cs, 12.39.Pn, 12.39.Jh}

\maketitle

\section{Introduction} \label{introduction}

The energy spectrum of $D_s$ mesons has remained poorly understood in comparison to charmonium spectrum. Currently, only 9 $D_s$ mesons are documented in the Particle Data Group (PDG). Among these, $D^{*}_{s0}(2317)$ stands out as an exceptionally intriguing meson due to its presence in the invariant mass spectrum of $D_s\pi$. This indicates a significant isospin violation within the $D^{*}_{s0}(2317)$ meson. This peculiar behavior can potentially be attributed to the fact that the lowest threshold channel in the $D_s$ system is the $DK$ energy at 2360 MeV, whereas the energy of $D^{*}_{s0}(2317)$ is notably lower than that of the $DK$ threshold. Consequently, $D^{*}_{s0}(2317)$ cannot directly decay into $DK$. One plausible explanation for this phenomenon is that $D^{*}_{s0}(2317)$ decays into $D_s\pi$ via $\pi$-$\eta$ mixing. Therefore, gaining a comprehensive understanding of $D^{*}_{s0}(2317)$ could contribute to our deeper understanding of Quantum Chromodynamics (QCD).

%Since $X(3872)$, which can't be accommodated into traditional quark model and near $D\bar{D}^*$ threshold, was reported by Belle Collaboration in 2003~\cite{01Choi:2003}, theorists and experimentalists had been showing great interest in exotic states. At beginning, Barnes \emph{et.al.} assigned
%$X(3872)$ into traditional charmonium~\cite{02Barnes:2004}, while T\"{o}rnqvist put forward that one pion exchange potential can make contribution to
%bind state of $X(3872)$~\cite{03Tornqvist:2004}, analogous to deuteron. Now, people tend to believe $X(3872)$ may be mixture of charmonium and mesonium~\cite{X3872_1,X3872_2,X3872_3,X3872_4}. Similarly, In 2003, BaBar Collaboration announced the other exotic state, $D^{*}_{s0}(2317)$, of which energy
%is less than prediction from traditional quark model and near $DK$ threshold~\cite{babar}. Subsequently, CLEO Collaboration confirmed the state and
%supplemented the other state $D_{s1}(2460)$ near the $D^*K$ threshold. Different from $X(3872)$ which mass is very close to $DD^*$ threshold, there are several
%dozen MeVs of energy difference between $D_{s0}(2317)$ and both traditional $P$-wave $c\bar{s}$ meson and $DK$ threshold~\cite{cleo}.

\begin{figure}[h]
 \vspace{-1.3cm}
\setlength {\abovecaptionskip} {-1.4cm}
  \centering
  \includegraphics[width=9cm,height=13cm]{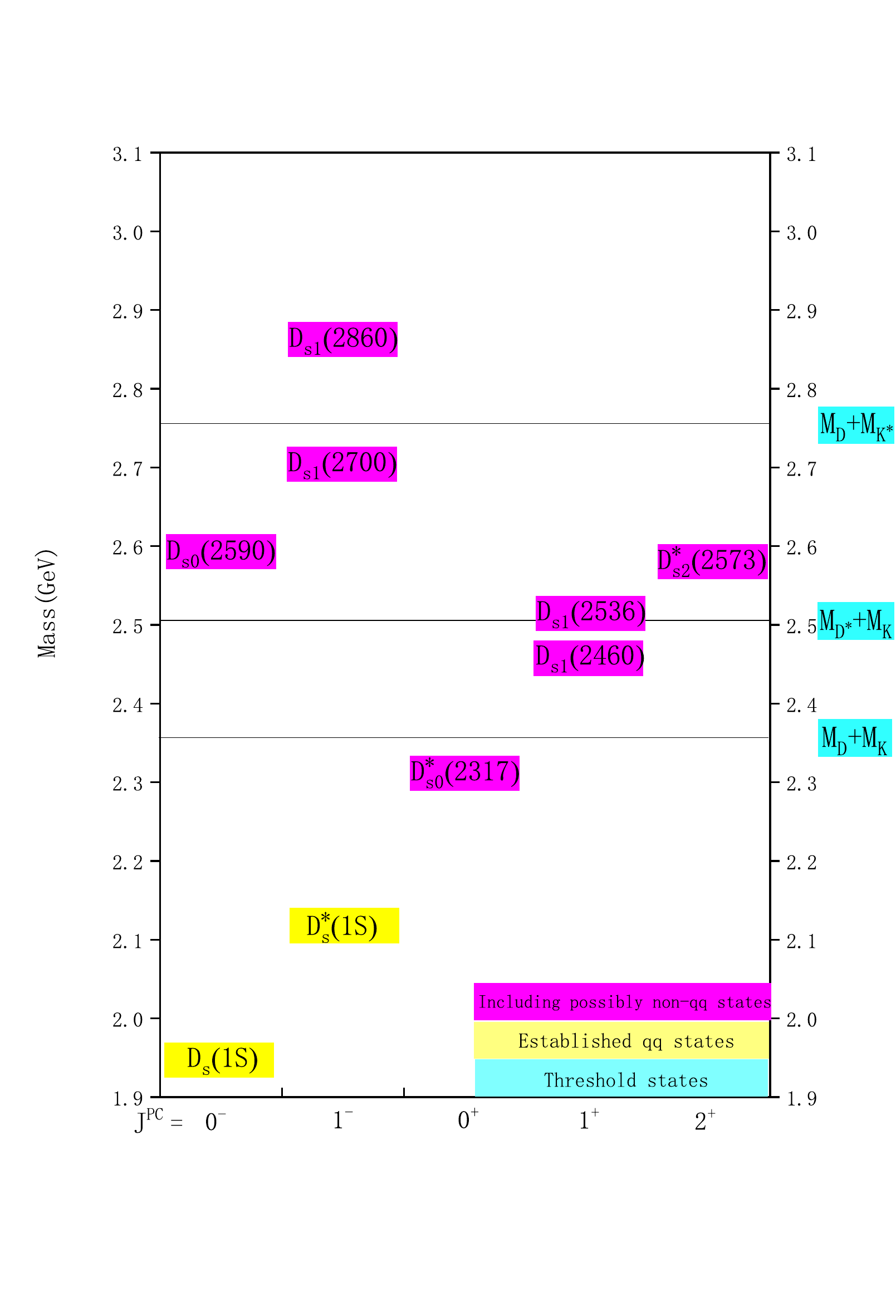}
  \caption{Energy spectrum of $D_s$ system.}
  \label{XYZ}
\end{figure}

There are a lot of work \cite{01,02,03,04,05,06,07,08,09,10,11,12,13,14,15,16,17,18,19,20,21,22,23,24,25,26,27,28,29,30,31,32,33,34,35,36,37,38,Kolomeitsev:2003ac} for answering
the question what's the nature structure of $D^*_{s0}(2317)$ and $D_{s1}(2460)$, and several relevant review articles \cite{Chen:2016spr,Guo:2017jvc} exist. 
Although the energy of ordinary meson $c\bar{s}$ has 100 MeV higher than $D^*_{s0}(2317)$ from Godfrey-Isgur model prediction~\cite{Godfrey:1985xj}, 
ordinary quark-antiquark picture was still invoked to explain the two exotic states~\cite{01,02,03,04,05,06}. Hadron mass is merely one property of particle, 
the mass difference between theoretical and experimental value can be
remedied by fine-tuning the model parameters. The transition probabilities which related to the internal structure of the state put more stringent test of the
model. For example, Godfrey \cite{01} assigned $^{3}P_0$ and $^{3}P_{1}$ quantum numbers to $D^*_{s0}(2317)$ and $D_{s1}(2460)$, and predicted the radiative decay
width of $D^*_{s0}\rightarrow D_s^{\pm} \gamma$, $D_{s1}\rightarrow D_s \gamma$. Because the mass of $D^*_{s0}(2317)$ is below strong decay threshold $DK$,
the radiative decay plays important role for understanding state $D^{*}_{s0}(2317)$. Due to two decay modes: $D_s^{\pm} \gamma$ and $D_s\pi$ are all discovered
experimentally, isospin violation in the transition is expected. Wang \emph{et~al}. \cite{03} studied the decay width of $D^*_{s0}(2317)$ and $D_{s1}(2460)$ in the chiral
quark model with $\pi-\eta$ mixing, the electromagnetic decay widths, but not pionic decay width, agreed with experimental data.
In addition, the $D^{*}_{s0}(2317)$ also can be ordinary $c\bar{s}$ in the frame work of thermal QCD sum rule~\cite{04}.

However, BESIII Collaboration obtained the absolute branching fraction $B(D^*_{s0}(2317)\rightarrow \pi^0 D_s)=1.00^{+0.00}_{-0.14}\pm 0.14$~\cite{bes3},
which differs from the expectation of the ordinary $c\bar{s}$ picture of $D^*_{s0}(2317)$~\cite{01}. Also the total decay width in the quark-antiquark picture
is too small to agree with experimental data. Then the assumption: $D^*_{s0}(2317)$ ($D_{s1}(2460)$) may be $DK$ ($D^*K$) molecular state,
is aroused~\cite{08,09,11,12,14,15,16}. Guo \emph{et al.} constructed an effective chiral Lagrangian including both electromagnetic decay width
and isospin violation decay width, and they got total width about $180\pm 110$ keV~\cite{12}, then updated the data to $133\pm 22$ keV~\cite{14}.
Faessler \emph{et al.} got the result $\Gamma(D^*_{s0}(2317) \rightarrow D_s\pi)=46.7$ keV by effective Lagrangian approach. Recently, Wu \emph{et al.}
found that the attractive interaction between $D$ and $K$ is strong enough to make a bound molecular state~\cite{19}. The tetraquark system was also invoked
to explain $D^*_{s0}(2317)$ ($D_{s1}(2460)$)~\cite{21,22,23,24,25}. Cheng \emph{et al.} treated $D^*_{s0}(2317)$ as $cq\bar{q}\bar{q}$ state, the mass and
decay width were obtained as $M=2320$ MeV and $\Gamma=11.3$ keV. By employing QCD sum rule, Zhang \emph{et al.} marked $D^*_{s0}(2317)$ as $0^+$
tetraquark to explain BESIII's results. Lattice QCD was also applied to investigate the corresponding four-quark state by calculating the four-quark
correlator, and they did not see a tetraquark mesonium in the $D^{*}_{s0}(2317)$ meson region~\cite{23}. The realistic quark model calculation also did not
support the molecular picture of $D_{s0}^*$ and $D_{s1}$~\cite{CPL26}. Our previous work showed that the interaction between $D$ and $K$ is repulsive while $D\bar{K}$ could be a possible partner of $D^{*}_{s0}(2317)$ due to "good di-quark"~\cite{Tan}.

Actually, more and more people are drawing their attention to mixture of $c\bar{s}$ and $DK$~\cite{27,28,29,30,31,32,33,34,35,36,37,38}.
In other words, the state $D^{*}_{s0}(2317)$ may be a $c\bar{s}$ state affected severely by $DK$ mesonium. Lattice QCD had a try to take
$S$-wave $DK$ threshold into consideration when studying $c\bar{s}$ system~\cite{30,31,32,33,34,36,37}, some of them claimed $c\bar{s}$ occupies
$70\%$ component in $D_{s0}^*(2317)$ state~\cite{36}.
What's more, Ortega \emph{et al.} utilized Isgur model with help of $^3P_0$ operator to study the state $D^*_{s0}(2317)$ as a mixture of
$c\bar{s}$ and $DK$ states, and got $66\%$ of $c\bar{s}$ component and $D_{s1}(2460)$ with $54\%$ $c\bar{s}$ component~\cite{35}.
Other decay channels also have contributions, Kolomeitsev shows the $D K$ and $D_s \eta$ coupling plays an important role in terms of the 
non-linear chiral SU(3) Lagrangian\cite{Kolomeitsev:2003ac}, and Torres \emph{et al.} obtained accurate $DK$ phase shifts and the position of the $D^{*}_{s0}(2317)$
in the framework of an $SU(4)$ extrapolation of the chiral unitary theory with couple of $DK$ and $D_s\eta$~\cite{32}.
Hence, in this work, we study states $D_{s0}^*(2317)$ and $D_{s1}(2460)$ in the unquenched quark model by considering all of possible channels
including $P$-wave and $D$-wave mesonia. In addition, the modified transition operator~\cite{qm5} which mixes the 2-quark and 4-quark
states is employed. The numeric results are obtained with the help of the high precision few-body method, Gaussian expansion method
(GEM)~\cite{29Hiyama:2003cu}.

The paper is organized as follows. In section II, the chiral quark model and GEM for solving the $q\bar q$ and $q\bar{q}$-$q\bar{q}$ systems
are presented. In Sec. III, we briefly introduce the unquenched quark model. The numerical results are given in Sec. IV. The last section is devoted
to the summary.

\section{Chiral quark model, wave function of $c\bar{q}q\bar{s}$ system} \label{GEM and chiral quark model}

\subsection{Chiral quark model}

The chiral quark model has been successfully applied to describe hadron spectra and hadron-hadron interactions. The detailed formulation of the model can be found in Refs. \cite{qm1,qm2,qm3,qm4,qm5}. In this work, we focus on the Hamiltonian for the four-quark system. Since we are primarily interested in the low-lying states, we consider only the central part of the quark-quark interactions for the four-quark system.
\begin{eqnarray}
H &=& \sum_{i=1}^nm_i+\sum_{i=1}^n(\frac{p_i^2}{2m_i}-T_{CM}) +  \sum_{i<j=1}^n [ V_{con}(r_{ij}) \nonumber\\
&+&V_{oge}(r_{ij}) + \sum_{\chi=\pi,\eta,K} V_{\chi}(r_{ij}) + V_{\sigma}(r_{ij}) ],
\end{eqnarray}

where \(m_i\) is the constituent mass of the \(i\)-th quark (or antiquark), and \(\mu\) is the reduced mass of two interacting quarks or quark-clusters. \(T_{\text{CM}}\) is the kinetic energy of the center-of-mass motion. For the two-quark system, the kinetic energy term \(\sum_{i=1}^n \left( \frac{p_i^2}{2m_i} - T_{\text{CM}} \right)\) degenerates into \(\frac{p_{12}^2}{2\mu_{12}}\), while the kinetic energy term of the tetraquark system can be written as:

\[
\frac{p_{12}^2}{2\mu_{12}} + \frac{p_{34}^2}{2\mu_{34}} + \frac{p_{12,34}^2}{2\mu_{12,34}}.
\]

The reduced masses are given by:

\[
\mu_{ij} = \frac{m_i m_j}{m_i + m_j}, \quad \text{for } ij = 12, 34,
\]
\[
\mu_{1234} = \frac{(m_1 + m_2)(m_3 + m_4)}{m_1 + m_2 + m_3 + m_4},
\]
\[
p_{ij} = \frac{m_j p_i - m_i p_j}{m_i + m_j},
\]
\[
p_{1234} = \frac{(m_3 + m_4) p_{12} - (m_1 + m_2) p_{34}}{m_1 + m_2 + m_3 + m_4}.
\]

The potential terms in the Hamiltonian include central, spin-orbit, and tensor forces. However, calculating the spin-orbit and tensor force contributions becomes complicated when dealing with four-quark systems. Therefore, we consider only the central force in the four-quark calculations, while spin-orbit coupling, tensor force, and central force are included in the two-quark calculations.

The first potential, \(V_{\text{con}}(r_{ij})\), represents the confining potential, which mimics the "confinement" property of QCD. This term includes both the central force \(V_{\text{con}}^{C}(r_{ij})\) and the spin-orbit force \(V_{\text{con}}^{SO}(r_{ij})\):

\begin{align}
\begin{split}
 \left \{
\begin{array}{ll}
    V_{con}^C(r_{ij}) &= ( -a_{c} r_{ij}^{2}-\Delta) \boldsymbol{\lambda}_i^c \cdot \boldsymbol{\lambda}_j^c\\
\\
    V_{con}^{SO}(r_{ij}) &= -\boldsymbol{\lambda}_i^c \cdot \boldsymbol{\lambda}_j^c \frac{a_c}{4m_i^2m_j^2}[ ((m_i^2+m_j^2)(1-2a_s) \\
    &+ 4m_im_j(1-a_s))(\vec{S}_{+}\cdot \vec{L})+\\
    &((m_j^2-m_i^2)(1-2a_s))(\vec{S}_{-}\cdot \vec{L}) ]\\
\end{array}
\right.
\end{split}
\end{align}

where \(\boldsymbol{\lambda}_i^c\) are the \(SU(3)\) color Gell-Mann matrices, and \(\vec{S}_{+}\) and \(\vec{S}_{-}\) refer to spin operators for the quark-antiquark pairs.

The second potential, \(V_{\text{oge}}(r_{ij})\), describes the one-gluon exchange interaction, which reflects the "asymptotic freedom" property of QCD. This term contains central, spin-orbit, and tensor force components:

\begin{align}
\begin{split}
 \left \{
\begin{array}{ll}
    V_{oge}^C(r_{ij}) &=\frac{\alpha_s}{4} \boldsymbol{\lambda}_i^c \cdot \boldsymbol{\lambda}_{j}^c
\left[\frac{1}{r_{ij}}-\frac{1}{6m_im_jr_0^2}\boldsymbol{\sigma}_i\cdot
\boldsymbol{\sigma}_j \frac{e^{-r_{ij}}/r_0(\mu_{ij})}{r_{ij}}\right]   \\
%&   \delta{(\boldsymbol{r}_{ij})}=\frac{e^{-r_{ij}/r_0(\mu_{ij})}}{4\pi r_{ij}r_0^2(\mu_{ij})}, \nonumber\\
    \\
    V_{oge}^{SO}(r_{ij}) &= -\frac{1}{16} \frac{\alpha_s\boldsymbol{\lambda}_i^c \cdot \boldsymbol{\lambda}_j^c}{4m_i^2m_j^2}[\frac{1}{r_{ij}^3}-\frac{e^{-r_{ij}/r_g(\mu)}}{r_{ij}^3}(1+\frac{r_{ij}}{r_g(\mu)})] \\
    &[ (m_i^2+m_j^2+4m_im_j)(\vec{S}_{+}\cdot \vec{L})\\
    &+(m_j^2-m_i^2)(\vec{S}_{-}\cdot \vec{L}) ]\\
V_{oge}^{T}(r_{ij}) &= -\frac{1}{16} \frac{\alpha_s\boldsymbol{\lambda}_i^c \cdot \boldsymbol{\lambda}_j^c}{4m_i^2m_j^2}[\frac{1}{r_{ij}^3}-\frac{e^{-r_{ij}/r_g(\mu)}}{r_{ij}}(\frac{1}{r_{ij}^2}\\
&+\frac{1}{3r^2_g(\mu)} +\frac{1}{r_{ij}r_g(\mu)})]S_{ij} \\
  %  V_{con}^{SO}(r_{ij}) = -\boldsymbol{\lambda}_i^c \cdot \boldsymbol{\lambda}_j^c \frac{a_c}{4m_i^2m_j^2}[ ((m_i^2+m_j^2)(1-2a_s)+4m_im_j(1-a_s))(\vector{S}_{+}\cdot \vector{L}) ]\\
\end{array}
\right.
\end{split}
\end{align}
where \(S_{ij}\) is the tensor operator for the quark pair.

The third potential, \(V_{\chi, s}\), represents the exchange of Goldstone bosons, which arises from the spontaneous breaking of chiral symmetry in QCD at low energies. This term includes contributions from pseudoscalar mesons (\(\pi, \eta, K\)) and scalar mesons (\(\sigma\)).

\begin{align}
\begin{split}
 \left \{
\begin{array}{ll}
V_{\chi}(r_{ij}) & =  v_{\pi}({{\bf r}_{ij}})\sum_{a=1}^{3} \lambda_i^a \lambda_j^a+v_{K}({{\bf r}_{ij}})\sum_{a=4}^{7}
	\lambda_i^a \lambda_j^a \\
&+v_{\eta}({{\bf r}_{ij}})[\cos\theta_{P}(\lambda_i^8 \lambda_j^8)-\sin\theta_{P}(\lambda_i^0 \lambda_j^0)] ,  \\
v_{\chi=\pi,K,\eta}^C & =  \frac{g^2_{ch}}{4\pi} \frac{m_{\chi}^2}{\Lambda_{\chi}^2-m_{\chi}^2} \frac{\Lambda^2_{\chi}}{\Lambda^2_{\chi}-m^2_{\chi}}m_{\chi}[ Y(m_{{\chi}}r_{ij})-\frac{\Lambda_{\chi}^3}{m_{\chi}^3}\\
&Y(\Lambda_{{\chi}}r_{ij})](\vec{\sigma}_i\cdot\vec{\sigma}_j),   \\
v_{{\chi}=\pi,K,\eta}^T & =  \frac{g^2_{ch}}{4\pi} \frac{m_{{\chi}}^2}{\Lambda_{\chi}^2-m_{\chi}^2} \frac{\Lambda^2_{\chi}}{\Lambda^2_{\chi}-m^2_{\chi}}m_{\chi}
	[ H(m_{{\chi}}r_{ij})-\frac{\Lambda_{\chi}^3}{m_{\chi}^3}\\
&H(\Lambda_{{\chi}}r_{ij})]S_{ij},   \\
\end{array}
\right.
\end{split}
\end{align}

\begin{align}
\begin{split}
 \left \{
\begin{array}{ll}
V_{\sigma}^C & =  -\frac{g^2_{ch}}{4\pi} \frac{\Lambda^2_{\sigma}}{\Lambda^2_{\sigma}-m^2_{\sigma}}m_{\sigma}
	\left[ Y(m_{{\sigma}}r_{ij})-\frac{\Lambda_{\sigma}}{m_s}Y(\Lambda_{{\sigma}}r_{ij})\right],   \\
V_{\sigma}^{SO} & =  -\frac{g^2_{ch}}{4\pi} \frac{\Lambda^2_{\sigma}}{\Lambda^2_{\sigma}-m^2_{\sigma}}\frac{m_{\sigma}^3}{2m_im_j}
	[ G(m_{\sigma}r_{ij}) \\
&-\frac{\Lambda_{\sigma}^3}{m_{\sigma}^3}G(\Lambda_{\sigma}r_{ij})]\vec{L}\cdot \vec{S},
\end{array}
\right.
\end{split}
\end{align}
where \(m_\pi, m_K, m_\eta\) are the masses of the pion, kaon, and eta mesons, and \(V_{\sigma}(r_{ij})\) represents the scalar meson exchange interaction.  $\boldsymbol{\lambda}$ are $SU(3)$ flavor Gell-Mann matrices, $m_{\chi(s)}$ is the masses of Goldstone bosons, $\Lambda_{\chi(s)}$ is the cut-offs, $g^2_{ch}/4\pi$ is the Goldstone-quark coupling constant.  The quark tensor operator is determined by $S_{ij}=3(\vec{\sigma}_i\cdot\hat{r}_{ij})(\vec{\sigma}_j\cdot\hat{r}_{ij})-(\vec{\sigma}_i\cdot\vec{\sigma}_j)$. Finally, $Y(x)$ is the standard Yukawa function defined by $Y(x)=e^{-x}/x$, $G(x)=(1+1/x)Y(x)/x$ and $H(x)=(1+3/x+3/x^2)Y(x)/x$.

The orbital wave function of the four-quark system consists of two sub-cluster orbital wave functions and the relative
motion wave function between two subclusters (1,3 denote quarks and 2,4 denote antiquarks),
\begin{equation}\label{spatialwavefunctions}
|R\rangle = \left[[\Psi_{l_1}({\bf r}_{12})\Psi_{l_2}({\bf
r}_{34})]_{l_{12}}\Psi_{L_r}({\bf r}_{1234}) \right]_{L}^{M_{L}},
\end{equation}
where the bracket "[~]" indicates angular momentum coupling, and $L$ is the total orbital angular momentum which
comes from the coupling of $L_r$, orbital angular momentum of relative motion, and $l_{12}$, which coupled by $l_1$ and $l_2$,
sub-cluster orbital angular momenta. In GEM, the radial part of the orbital wave function is expanded by a set of Gaussians:
\begin{equation}
\label{radialpart}
\Psi(\mathbf{r})  = \sum_{n=1}^{n_{\rm max}} c_{n} N_{nl}r^{l} e^{-\nu_{n}r^2}Y_{lm}(\hat{\mathbf{r}}),
\end{equation}

where $N_{nl}$ are normalization constants,
\begin{align}
N_{nl}=\left[\frac{2^{l+2}(2\nu_{n})^{l+\frac{3}{2}}}{\sqrt{\pi}(2l+1)}
\right]^\frac{1}{2}.
\end{align}
$c_n$ are the variational parameters, which are determined dynamically. The Gaussian size parameters are chosen according
to the following geometric progression
\begin{equation}\label{gaussiansize}
\nu_{n}=\frac{1}{r^2_n}, \quad r_n=r_1a^{n-1}, \quad
a=\left(\frac{r_{n_{\rm max}}}{r_1}\right)^{\frac{1}{n_{\rm max}-1}}.
\end{equation}
This procedure enables optimization of the using of Gaussians, as small as possible Gaussians are used.

All the parameters are determined by fitting the meson spectrum, from light to heavy, taking into account only a quark-antiquark component.
%They are shown in Table~\ref{modelparameters}. The calculated masses of the mesons involved in the present work are shown in
%Table~\ref{mesonmass}. For the quark-antiquark calculation, the spin-orbit interactions and tensor interactions are also taken into account,
%The detailed form of these interactions can be found in Ref.~\cite{qm1}.
%
%From Table~\ref{mesonmass}, one can see that the calculated masses of mesons agree with the experimental data well, the two $P$-wave mesons,
%$D_{s0}^*$ and $D_{s1}$, the masses are 2336 MeV and 2433 MeV, which are very close to that of $D^{*}_{s0}(2317)$ and $D_{s1}(2460)$.
%However, the large isospin-violeation decay width disfavor the $c\bar{s}$ assignment~\cite{bes3}. In the following, we study these two states
%in the unquenched quark model.

\begin{table}[t]
\begin{center}
\caption{Quark model parameters ($m_{\pi}=0.7$ fm, $m_{\sigma}=3.42$ fm, $m_{\eta}=2.77$ fm, $m_{K}=2.51$ fm).\label{modelparameters}}
\begin{tabular}{cccc}
\hline\hline\noalign{\smallskip}
Quark masses   &$m_u=m_d$ (MeV)     &313  \\
               &$m_{s}$ (MeV)         &536  \\
               &$m_{c}$ (MeV)         &1728 \\
               &$m_{b}$ (MeV)         &5112 \\
\hline
Goldstone bosons   %&$m_{\pi}(fm^{-1})$     &0.70  \\
%                   &$m_{\sigma}(fm^{-1})$     &3.42  \\
%                   &$m_{\eta}(fm^{-1})$     &2.77  \\
%                   &$m_{K}(fm^{-1})$     &2.51  \\
                   &$\Lambda_{\pi}=\Lambda_{\sigma}$(fm$^{-1}$)     &4.2  \\
                   &$\Lambda_{\eta}=\Lambda_{K}$(fm$^{-1}$)      &5.2  \\
                   &$g_{ch}^2/(4\pi)$                &0.54  \\
                   &$\theta_p(^\circ)$                &-15 \\
\hline
Confinement             &$a_{c}$ (MeV/fm$^2$)     &101 \\
%                        &$a_{s}$     &0.777 \\
                   &$\Delta$ (MeV)       &-78.3 \\
%                   &$\mu_{c}$ (MeV)       &0.7 \\
\hline
OGE                 & $\alpha_{0}$        &3.67 \\
                   &$\Lambda_{0} $(fm$^{-1}$) &0.033 \\
                  &$\mu_0$ (MeV)    &36.976 \\
                   &$\hat{r}_0$ (MeV)    &28.17 \\
                   & $a_{s}$        &0.77 \\
\hline\hline
\end{tabular}
\end{center}
\end{table}

\subsection{The wave function of $c\bar{q}q\bar{s}$ system}

The wave function of the $c\bar{q}q\bar{s}$ system should be the direct product of the orbital wave function ( $|R\rangle$ ) of $L=0$, the spin wave function ($|S_{i}\rangle$) of the total $S=0$, the flavor wave function ($|F_j\rangle$) of $I=0$, and the color wave function ($|C_k \rangle$), according to the $J^P$ of $D_{s0}^{*}(2317)$ = $0^+$, which can be written as:
\begin{equation}
|\Psi^{i,j,k} \rangle={\cal A}    |R\rangle  |S_{i}\rangle  |F_j\rangle |C_k \rangle
\end{equation}
where $\cal A$ is the system's antisymmetry operator, which ensures that when identical particles exchange, the total wave functions will be antisymmetric. The ${\cal A}=1$ because there are no identical particles in the $c\bar{q}q\bar{s}$ tetraquark system. Finally, using the Rayleigh-Ritz variational principle, we solve the following Schr\"{o}dinger equation to get the system's eigenenergies.

\subsubsection{flavor wave function\label{sec_flavor}}

We have three flavor wave functions of the system,
\begin{eqnarray}\label{sec}
|F_{1}\rangle&=\frac{1}{\sqrt{2}}( c_1\bar{u}_2u_3\bar{s}_4+c_1\bar{d}_2d_3\bar{s}_4),\\
|F_{2}\rangle&=\frac{1}{\sqrt{2}}( c_1\bar{s}_4u_3\bar{u}_2+c_1\bar{s}_4d_3\bar{d}_2),\\
|F_{3}\rangle&=\frac{1}{\sqrt{2}}( c_1u_3\bar{u}_2\bar{s}_4+c_1d_3\bar{d}_2\bar{s}_4)
\end{eqnarray}
$|F_{1}\rangle, |F_{2}\rangle$ is for meson-meson structure, and $|F_{3}\rangle$ is for diquark-antidiquark structure.

\begin{figure}[htbp]
	\centering
	\begin{minipage}{0.49\linewidth}
		\centering
		\includegraphics[width=0.9\linewidth]{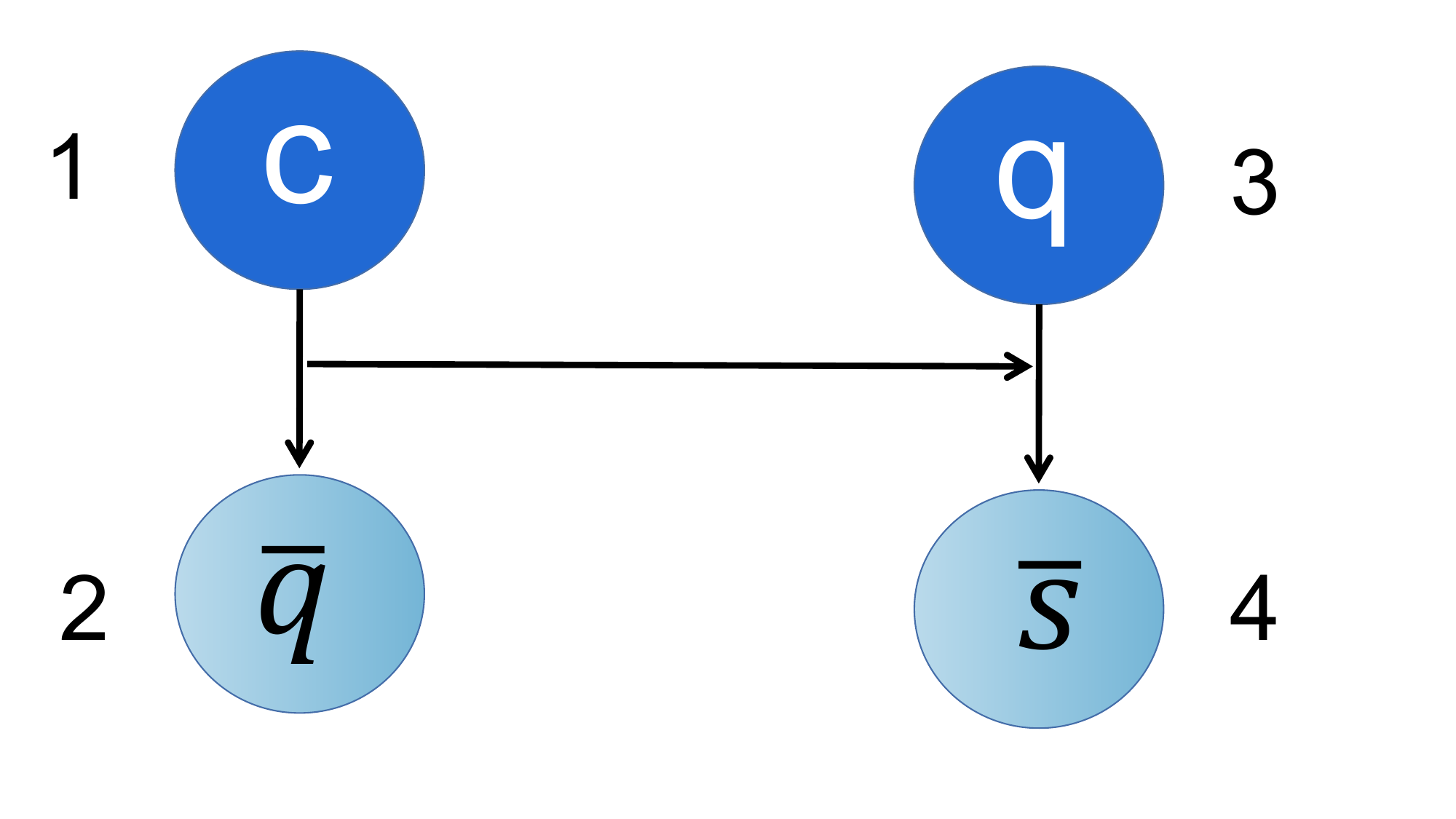}
		\caption{$c\bar{q}$-$q\bar{s}$ structure}
		\label{M1}%ÎÄÖÐÒýÓøÃͼƬ´úºÅ
	\end{minipage}
	\begin{minipage}{0.49\linewidth}
		\centering
		\includegraphics[width=0.9\linewidth]{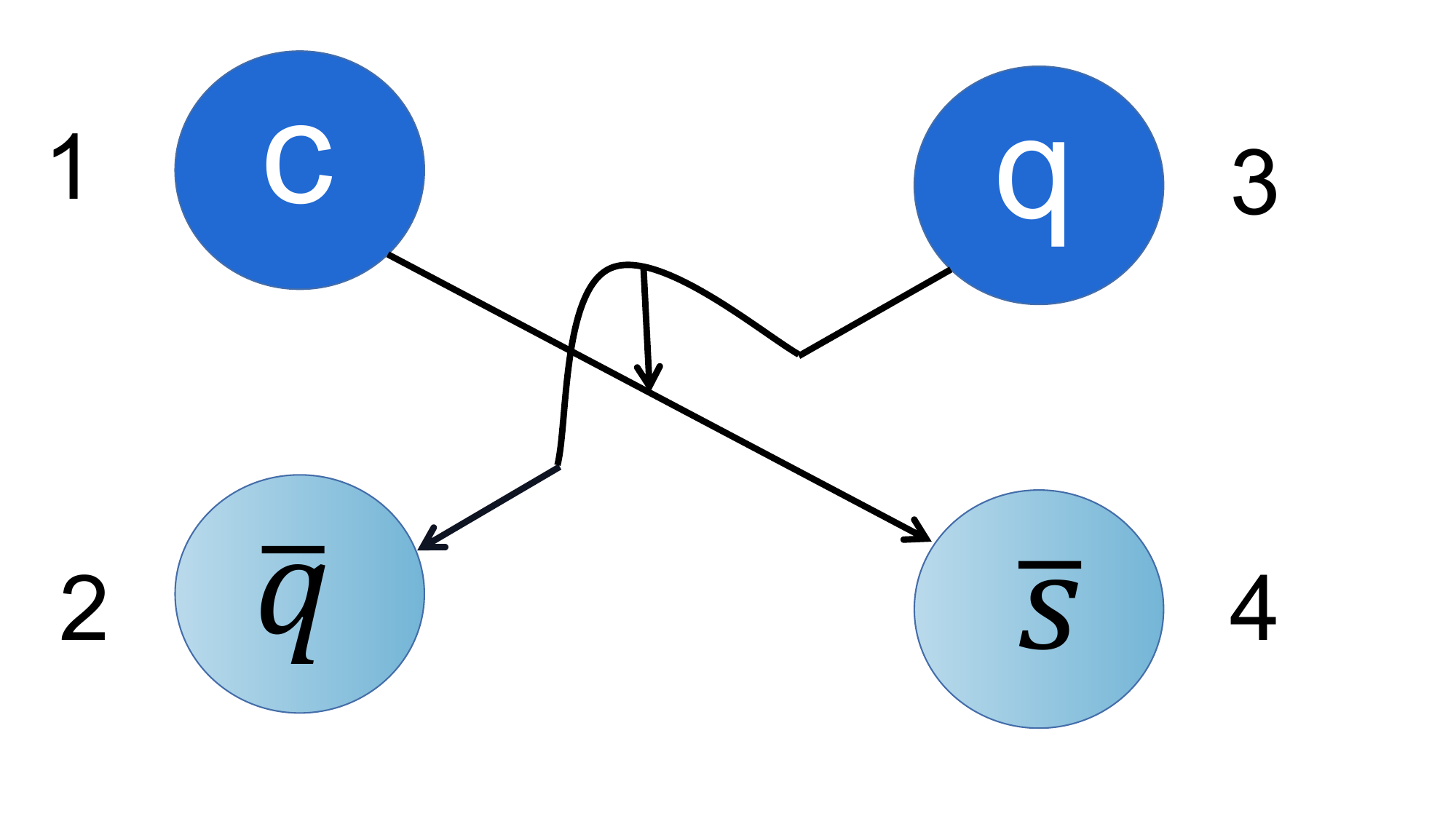}
		\caption{$c\bar{s}$-$q\bar{q}$ structure}
		\label{M2}%ÎÄÖÐÒýÓøÃͼƬ´úºÅ
	\end{minipage}
	%\qquad
	%ÈÃͼƬ»»ÐУ¬
	
	\begin{minipage}{0.49\linewidth}
		\centering
		\includegraphics[width=0.9\linewidth]{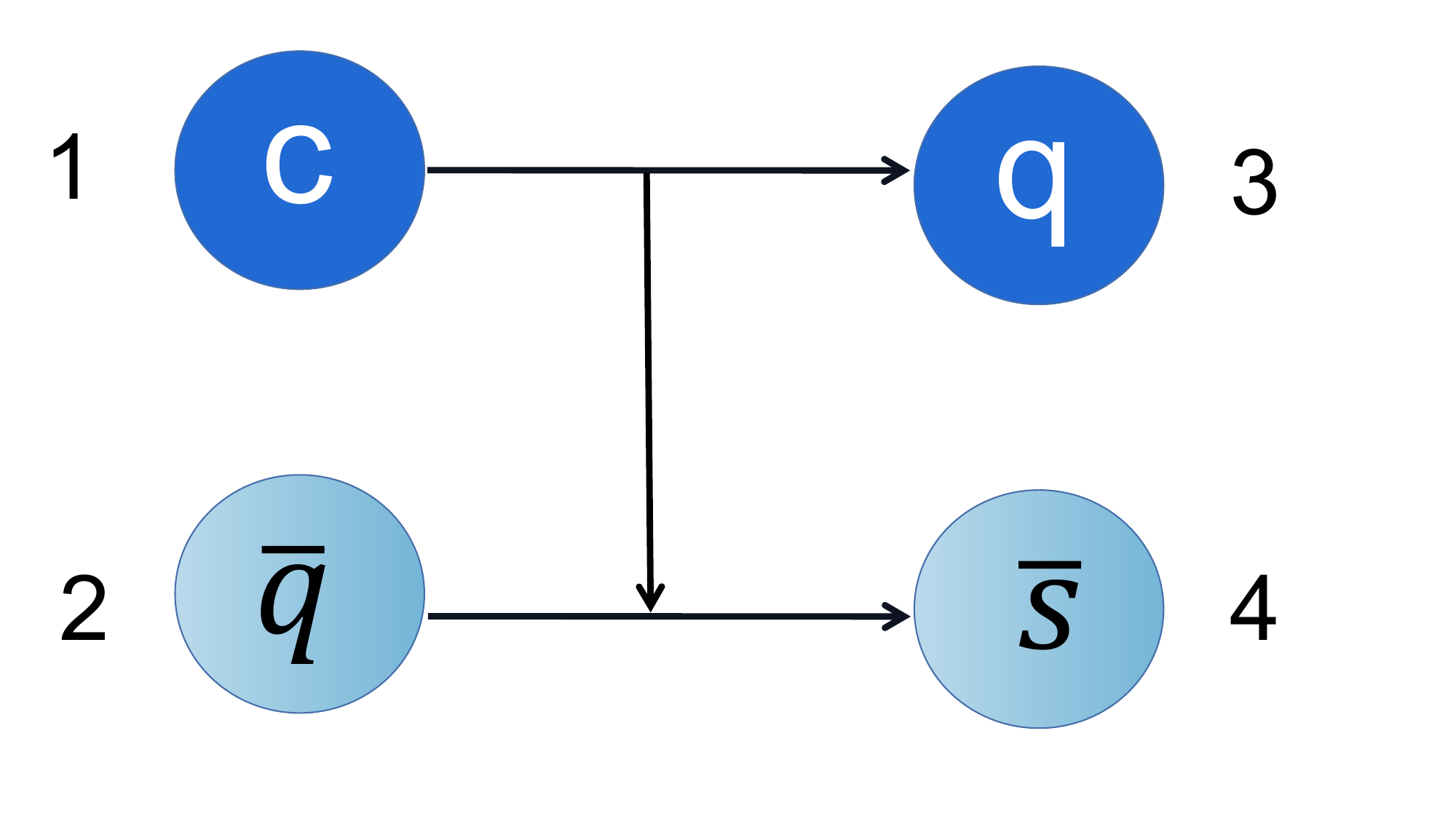}
		\caption{$cq$-$\bar{q}\bar{s}$ structure}
		\label{D1}%ÎÄÖÐÒýÓøÃͼƬ´úºÅ
	\end{minipage}
\end{figure}

\subsubsection{spin wave function}
Because of no difference between spin of quark and antiquark, the meson-meson structure has the same spin wave function as
the diquark-antidiquark structure. The spin wave functions of the sub-cluster are shown below.
\begin{align*}
&\chi_{11}^{\sigma}=\alpha\alpha,~~
\chi_{10}^{\sigma}=\frac{1}{\sqrt{2}}(\alpha\beta+\beta\alpha),~~
\chi_{1-1}^{\sigma}=\beta\beta,\nonumber \\
&\chi_{00}^{\sigma}=\frac{1}{\sqrt{2}}(\alpha\beta-\beta\alpha),
\end{align*}
Coupling the spin wave functions of two sub-clusters by Clebsch-Gordan coefficients, total spin wave function can be written below,

\begin{align}
&|S_{1}\rangle=\chi_{00}^{\sigma} \chi_{00}^{\sigma}=  \sqrt{\frac{1}{4}}(\alpha_1\beta_2\alpha_3\beta_4-\alpha_1\beta_2\beta_3\alpha_4-\beta_1\alpha_2\alpha_3\beta_4 \nonumber\\
&+\beta_1\alpha_2\beta_3\alpha_4 ),\\
&|S_{2}\rangle= \sqrt{\frac{1}{3} } ( \chi_{1-1}^{\sigma} \chi_{11}^{\sigma}+\chi_{11}^{\sigma} \chi_{1-1}^{\sigma}-\chi_{00}^{\sigma} \chi_{00}^{\sigma} ) \nonumber\\ &=\sqrt{\frac{1}{12}}(2\alpha_1\alpha_2\beta_3\beta_4+2\beta_1\beta_2\alpha_3\alpha_4-\alpha_1\beta_2\alpha_3\beta_4 \nonumber\\
&-\alpha_1\beta_2\beta_3\alpha_4-\beta_1\alpha_2\alpha_3\beta_4-\beta_1\alpha_2\beta_3\alpha_4 ).
\end{align}

The total spin wave function is denoted by $|S_{i}\rangle$, $i$ is the index of the functions. The second spin wave function $|S_{2}\rangle$ demonstrates that the wave functions of two subclusters
with spin 1 are coupled to a total spin of 0, while the first spin wave function $|S_{1}\rangle$ shows that the spin quantum numbers of both subclusters are both 0. 
Due to the spin symmetry of the Hamiltonian which can distinguish the third component of the spin quantum number, we set the third component of the spin to be zero for simplicity.

\subsubsection{color wave function}
The colorless tetraquark system has four color wave functions, two for meson-meson structure, $1\otimes1$ ($C_1$), $8\otimes8$ ($C_2$),
and two for diquark-antidiquark structure, $\bar{3}\otimes 3$ ($C_3$) and $6\otimes \bar{6}$ ($C_4$).

\begin{align}
&|C_{1}\rangle =  \sqrt{\frac{1}{9}} ( r_1\bar{r}_2r_3\bar{r}_4+r_1\bar{r}_2g_3\bar{g}_4+r_1\bar{r}_2b_3\bar{b}_4+g_1\bar{g}_2r_3\bar{r}_4 \nonumber\\
& +g_1\bar{g}_2g_3\bar{g}_4 +g_1\bar{g}_2b_3\bar{b}_4+b_1\bar{b}_2r_3\bar{r}_4  +b_1\bar{b}_2g_3\bar{g}_4+b_1\bar{b}_2b_3\bar{b}_4)  \nonumber\\
&|C_{2}\rangle = \sqrt{\frac{1}{72}}(3r_1\bar{b}_2b_3\bar{r}_4+3r_1\bar{g}_2g_3\bar{r}_4+3g_1\bar{b}_2b_3\bar{g}_4                   \nonumber\\
&+3b_1\bar{g}_2g_3\bar{b}_4+3g_1\bar{r}_2r_3\bar{g}_4+3b_1\bar{r}_2r_3\bar{b}_4+2r_1\bar{r}_2r_3\bar{r}_4 \nonumber\\
&+2g_1\bar{g}_2g_3\bar{g}_4+2b_1\bar{b}_2b_3\bar{b}_4 -r_1\bar{r}_2g_3\bar{g}_4-g_1\bar{g}_2r_3\bar{r}_4\nonumber\\
&-b_1\bar{b}_2g_3\bar{g}_4-b_1\bar{b}_2r_3\bar{r}_4-g_1\bar{g}_2b_3\bar{b}_4-r_1\bar{r}_2b_3\bar{b}_4). \nonumber\\
\end{align}
\begin{align}
&|C_{3}\rangle = \sqrt{\frac{1}{12}}(r_1g_3\bar{r}_2\bar{g}_4-r_1g_3\bar{g}_2\bar{r}_4+g_1r_3\bar{g}_2\bar{r}_4-g_1r_3\bar{r}_2\bar{g}_4 \nonumber \\
&+r_1b_3\bar{r}_2\bar{b}_4-r_1b_3\bar{b}_2\bar{r}_4+b_1r_3\bar{b}_2\bar{r}_4-b_1r_3\bar{r}_2\bar{b}_4+g_1b_3\bar{g}_2\bar{b}_4 \nonumber \\
&-g_1b_3\bar{b}_2\bar{g}_4+b_1g_3\bar{b}_2\bar{g}_4-b_1g_3\bar{g}_2\bar{b}_4). \nonumber \\
&|C_{4}\rangle = \sqrt{\frac{1}{24}}(2r_1r_3\bar{r}_2\bar{r}_4+2g_1g_3\bar{g}_2\bar{g}_4+2b_1b_3\bar{b}_2\bar{b}_4+r_1g_3\bar{r}_2\bar{g}_4 \nonumber \\
&    +r_1g_3\bar{g}_2\bar{r}_4 +g_1r_3\bar{g}_2\bar{r}_4+g_1r_3\bar{r}_2\bar{g}_4+r_1b_3\bar{r}_2\bar{b}_4 +r_1b_3\bar{b}_2\bar{r}_4\nonumber \\
&+b_1r_3\bar{b}_2\bar{r}_4 +b_1r_3\bar{r}_2\bar{b}_4+g_1b_3\bar{g}_2\bar{b}_4+g_1b_3\bar{b}_2\bar{g}_4+b_1g_3\bar{b}_2\bar{g}_4\nonumber \\
&+b_1g_3\bar{g}_2\bar{b}_4).\nonumber \\
\end{align}

\section{Unquenched quark model} \label{T operator}
% \subsection{method of mass shift and component}
In unquenched quark model, the high Fock components are taken into account. It is expected that the na\"{i}ve quark model is a good zeroth order
approximation for hadron states, so we consider only four-quark components for mesons,
\begin{eqnarray}
 \Psi=c_2\Psi_{2q}+\sum_{i=1}^{N}c_{4i}\Psi^i_{4q} \,,
\end{eqnarray}
where $\Psi_{2q}$ and $\Psi_{4q}^i$ are the wave functions of the two- and four-quark components, respectively, and N is the total number of
four-quark channels. The expansion coefficients are determined by solving the Schr\"{o}dinger equation,
\begin{eqnarray}
  H\Psi = E\Psi .
\end{eqnarray}
In the nonrelativistic quark model, the number of particles is conserved. So to write down the Hamiltonian of the nonrelativistic unquenched quark model
in a usual way is not possible. Generally the prescription is given to the following Hamiltonian,
\begin{eqnarray}
  H=H_{2q}+H_{4q}+T_{24} .
\end{eqnarray}
$H_{2q}$ acts only on the wave function, $\Psi_{2q}$; $H_{4q}$ acts only on the wave functions $\Psi_{4q}^i$; and $T_{24}$ couples the two- and four-quark components.

The $^3P_0$ model (quark pair creation model) was originally introduced by Micu \cite{15Micu:1969} and further developed by Le
Yaouanc, Ackleh and Roberts \textit{et al}.~\cite{16Yaouanc:1973,17Yaouanc:1974,22ACKleh:1996,23Roberts:1992}. It can be applied to the OZI
rule allowed two-body strong decays of a hadron~\cite{15Micu:1969,22ACKleh:1996,24Capstic:1986,25Capstic:1994,26page:1995}. The
transition operator in the model is,
\begin{eqnarray} \label{T0}
  &&T_1=-3~\gamma\sum_m\langle 1m1-m|00\rangle\int
  d\mathbf{p}_3d\mathbf{p}_4\delta^3(\mathbf{p}_3+\mathbf{p}_4)\nonumber\\
  &&~~~~\times{\cal{Y}}^m_1(\frac{\mathbf{p}_3-\mathbf{p}_4}{2})
  \chi^{34}_{1-m}\phi^{34}_0\omega^{34}_0b^\dagger_3(\mathbf{p}_3)d^\dagger_4(\mathbf{p}_4),
\end{eqnarray}
where $\gamma$ represents the probability of the quark-antiquark pair with momentum $\mathbf{p}_3$ and $\mathbf{p}_4$
created from the vacuum. Because the intrinsic parity of the
antiquark is negative, the created quark-antiquark pair must be in the state $^{2S+1}L_{J}={}^3P_0$.
$\phi^{34}_{0}$ and $\omega^{34}_0$ are flavor and color singlet states, respectively (the quark and the
antiquark in the original meson are indexed by 1 and 2). The S-matrix element for the
process $A \rightarrow B + C$ is written as
\begin{equation}
 \langle BC|T|A\rangle=\delta^3(\mathbf{P}_A-\mathbf{P}_B-\mathbf{P}_C){\cal{M}}^{M_{J_A}M_{J_B}M_{J_C}},
\end{equation}
where $\mathbf{P}_B$ and $\mathbf{P}_C$ are the momenta of B and C mesons in the final state, and
satisfy $\mathbf{P}_A = \mathbf{P}_B + \mathbf{P}_C = 0$ in the center-of-mass frame of meson A.
${\cal{M}}^{M_{J_A}M_{J_B}M_{J_C}}$ is the helicity amplitude of the process $A \rightarrow B + C$.

Using the above transition operator in the unquenched quark model to calculate the mass shifts of light mesons
duo to the coupling of four-quark components, one obtained too large values~\cite{qm5}. To deal with this issue,
a modified transition operator was proposed~\cite{qm5}. It reads,
\begin{eqnarray} \label{T2}
   T_2&=&-3\gamma\sum_{m}\langle 1m1-m|00\rangle\int
   d\mathbf{r_3}d\mathbf{r_4}\left(\frac{1}{2\pi}\right)^{\frac{3}{2}}ir2^{-\frac{5}{2}}f^{-5}
   \nonumber \\
   &&Y_{1m}(\hat{\mathbf{r}})e^{-\frac{r^2}{4f^2}}e^{-\frac{R_{AV}^2}{R_0^2}}\chi_{1-m}^{34}\phi_{0}^{34}
   \omega_{0}^{34}b_3^{\dagger}(\mathbf{r_3})d_4^{\dagger}(\mathbf{r_4}).
\end{eqnarray}
Here, ``A" stands for the bare meson and ``V" denotes the antiquark-quark pair created in the vacuum, hence, $R_{AV}$ is the relative distance between
the source particle A and quark-antiquark pair V in the vacuum.
\begin{eqnarray*}
  \mathbf{R}_{AV}&=&\mathbf{R}_A-\mathbf{R}_V;
\end{eqnarray*}
\begin{eqnarray*}
   \mathbf{R}_A&=&\frac{m_1\mathbf{r_1}+m_2\mathbf{r_2}}{m_1+m_2};
\end{eqnarray*}
\begin{eqnarray*}
   \mathbf{R}_V&=&\frac{m_3\mathbf{r_3}+m_4\mathbf{r_4}}{m_3+m_4}=\frac{\mathbf{r_3}+\mathbf{r_4}}{2}~~(m_3=m_4).
\end{eqnarray*}
The damping factors in the modified transition operator mainly consider the effect of quark-antiquark created in the vacuum. The first damping factor
$e^{-r^2/(4f^2)}$ (the corresponding expression in the momentum space is $e^{-f^2 p^2}$) suppresses the creation of quark-antiquark
with high energy, and the second damping factor $e^{-R^2_{AV}/R^2_0}$ takes account of the fact that the created quark-antiquark pair
should not be far away from the original meson.

The parameter $\gamma$ is generally determined by an overall fitting of the strong decay width of hadrons. The relation between the strength
$\gamma_{u,d}$ for $u\bar{u}$, $d\bar{d}$ creation and the strength $\gamma_{s}$ for $s\bar{s}$ creation is $\gamma_{s}=\gamma_{u}/\sqrt{3}$~\cite{28PLB}.
In this work, the parameters in the modified transition operator are taken from Ref.~\cite{qm5},
\begin{eqnarray*}
  \gamma_{u,d}=32.2,~~f=0.7~\mbox{fm$^2$},~~R_{0}=1.0~\mbox{fm$^2$}.
\end{eqnarray*}

\section{Results and discussions}

To investigate whether $D^{*}_{s0}(2317)$ may exist as a two-quark state, we first compute the $D_s$ spectrum in the two-quark framework in this section. We then perform a systematic calculation of the $c\bar{q}q\bar{s}$ tetraquark system with $0^+$ using the GEM  approach to investigate the plausibility of $D^{*}_{s0}(2317)$ as a molecular state. The mixture of two-quark $c\bar{s}$ and four-quark $DK$ molecules is then calculated using the $^3P_0$ model.

\subsection{Two-quark calculation }

In Table \ref{mesonmass}, we list our calculation results. In the second column, the values before the "/" are the results from calculations considering only the central force, while those after the "/" include the spin-orbit coupling interaction.

Most of the model calculations \cite{Zeng:1994vj}\cite{Godfrey:1985xj} currently suggest that the energy of the $c\bar{s}$ structure of $D_{s0}^{*}(2317)$ is around 2.4 GeV, which implies that $D_{s0}^{*}(2317)$ is unlikely to be a two-quark state with the $c\bar{s}$ structure. According to our calculations, if we consider only the central force, the energy of the $c\bar{s}$ structure of $D_{s0}^{*}(2317)$ is 2.48 GeV, which is consistent with the results from other models. However, when we introduce the spin-orbit coupling interaction into the calculations, we find that the energy of the $c\bar{s}$ structure of $D_{s0}^{*}(2317)$ is 2.33 GeV, while the energy of the $c\bar{s}$ structure of $D_{s1}^{\prime}(2460)$ is 2.43 GeV, both of which are more in line with the experimental values. This shows that the energy of the $c\bar{s}$ structure of $D_{s0}^{*}(2317)$ is sensitive to the spin-orbit coupling interaction, which is also obtained 
in the work by Vijande et al. \cite{Vijande:2004he}. Vijande calculated the energy of the $c\bar{s}$ structure of $D_{s0}^{*}(2317)$ to be 2.46 GeV, and after adjusting the parameter $a_s$, this energy can reach 2.32 GeV, which is very close to the experimental value.

In fact, there is also a lattice QCD study \cite{Dong:2009wk} that found the energy of the $c\bar{s}$ structure of $D_{s0}^{*}(2317)$ to be 2.3 GeV, which supports the view that $D_{s0}^{*}(2317)$ could be a two-quark state with a $c\bar{s}$ structure. Therefore, from the energy perspective, the two-quark $c\bar{s}$ structure can explain the $D_{s0}^{*}(2317)$.

\begin{table}[]
\caption{ \label{mesonmass}  Numerical results for different models (unit: MeV).\label{mesons}}
\begin{tabular}{ccccccc}
\hline\hline\noalign{\smallskip}
    Meson             &  Our work   &     GI\cite{Godfrey:1985xj}     &   CQM\cite{Vijande:2004he} &   LQCD\cite{Dong:2009wk} &   HQET \cite{Zeng:1994vj} & EXP. \\ \hline
    $D$               &    1863     &     1880   &  1981 &   -    &   -    & 1867.7\\
    $D^*$             &    1981     &     2040   &  2112 &   -    &   -    &  2008.9\\
    $K$               &     494     &     470    &   496 &   -    &   -    & 495.0\\
    $K^*$             &     914     &     900    &   910 &   -    &   -    & 892.0\\
    $\eta^{\prime}$   &     824     &     960    &   956 &   -    &   -    & 957.8\\
    $\phi$            &    1016     &     1020   &  1020 &   -    &   -    & 1019.0\\
    $D_s$             &    1953     &     1980   &  1981 &   1969 &   1969 & 1968.0\\
    $D_s^*$           &    2080     &     2130   &  2112 &   2121 &   2111 & 2112.0\\
    $D_{s1}^{\prime}$ &    2480/2433&     2530   &  2482 &   -    &   2521 & 2460.0\\
    $D_{s0}$          &    2479/2336&     2480   &  2317 &   2304 &   2387 & 2317.0\\
\hline\hline
\end{tabular}
\end{table}

\subsection{Tetraquark calculation }

\begin{table}[]
\caption{ \label{tetraquark} The energy of DK system (unit: MeV)}
\begin{tabular}{ccccccccccccccc}
\hline\hline\noalign{\smallskip}
$|F_j\rangle$ &$|S_i\rangle$ & $|C_k\rangle$ &Channel&~~$E$~~&$E^{Theo}_{th}$&$E_{B}$&~~$E^{Exp}_{th}$\\
\hline
$|F_1\rangle$ &$|S_1\rangle$ & $|C_1\rangle$  & $DK$                                             &  2359&  2357&0  &  2362.7  \\
$|F_1\rangle$ &$|S_1\rangle$ & $|C_2\rangle$  & $[D]_8[K]_8$                                     &  3155&  $-$   &$-$  &  $-$  \\
$|F_1\rangle$ &$|S_2\rangle$ & $|C_1\rangle$  & $D^*K^*$                                         &  2625&  2984&0  &  2900.9 \\
$|F_1\rangle$ &$|S_2\rangle$ & $|C_2\rangle$  & $[D^*]_8[K^*]_8$                                 &  3078&  $-$   &$-$&  $-$    \\
$|F_2\rangle$ &$|S_1\rangle$ & $|C_1\rangle$  & $D_s \eta$                                       &  2625&  2623&$0$&  2515.7  \\
$|F_2\rangle$ &$|S_1\rangle$ & $|C_2\rangle$  & $[D_s]_8[\eta]_8$                                &  3218&  $-$   &$-$&  $-$     \\
$|F_2\rangle$ &$|S_2\rangle$ & $|C_1\rangle$  & $D_s^* \omega$                                     &  2785&  2782&$0$&  2894.0 \\
$|F_2\rangle$ &$|S_2\rangle$ & $|C_2\rangle$  & $[D_s^*]_8[\phi]_8$                              &  2928&  $-$   &$-$&  $-$     \\
$|F_3\rangle$ &$|S_1\rangle$ & $|C_3\rangle$  & $[cq]_3^0[\bar{q}\bar{s}]_{\bar{3}}^0$           &  3003&  $-$   &$-$&  $-$     \\
$|F_3\rangle$ &$|S_1\rangle$ & $|C_4\rangle$  & $[cq]_3^1[\bar{q}\bar{s}]_{\bar{3}}^1$           &  3163&  $-$   &$-$ &  $-$    \\
$|F_3\rangle$ &$|S_2\rangle$ & $|C_3\rangle$  & $[cq]_6^0[\bar{q}\bar{s}]_{\bar{6}}^0$           &  3181&  $-$   &$-$  &  $-$    \\
$|F_3\rangle$ &$|S_2\rangle$ & $|C_4\rangle$  & $[cq]_6^1[\bar{q}\bar{s}]_{\bar{6}}^1$           &  3060&  $-$   &$-$  &  $-$    \\
\multicolumn{4}{l}{Complete coupled-all-channels:}   &2357&2357&0\\
\hline\hline
\end{tabular}
\end{table}

%In the tetraquark system, there are a total of four molecular channels with a color-singlet state: \(DK\), \(D^{*}K^{*}\), \(D_s \eta\), and \(D_s^{*}\omega\), along with their corresponding four color-octet states. Their energy range lies between 2.3 and 2.7 GeV. According to our calculations (as shown in Table \ref{tetraquark}), their binding energies (\(E_B\)) are zero, indicating that they are all scattering states.
%

In the four-quark framework of $D_s(2317)$, there are a total of four color-singlet and four color-octet molecular states, as well as four diquark structures. Among these, the molecular states belonging to the $c\bar{q}$-$q\bar{s}$ structure are $DK$, $[D]_8[K]_8$, $D^*K^*$, and $[D^*]_8[K^*]_8$, while those belonging to the $c\bar{s}$-$q\bar{q}$ structure are $D_s \eta$, $[D_s]_8[\eta]_8$, and $D_s^* \omega$, $[D_s^*]_8[\omega]_8$. The diquark structures are $[cq]_3^0[\bar{q}\bar{s}]_{\bar{3}}^0$, $[cq]_3^1[\bar{q}\bar{s}]_{\bar{3}}^1$, $[cq]_6^0[\bar{q}\bar{s}]_{\bar{6}}^0$, and $[cq]_6^1[\bar{q}\bar{s}]_{\bar{6}}^1$. Our calculations (as shown in Table \ref{tetraquark}) show that the energies of the color-octet and diquark structures are relatively high, in the range of 2.9 to 3.1 GeV, and above the corresponding thresholds, indicating weak attraction between these physical channels. The color-singlet states $DK$, $D^*K^*$, $D_s\eta$, and $D_s^* \omega$ are not bound states. Afterward, we performed channel coupling calculations, and the results indicate that the channel coupling effect lowers the energy of the lowest-energy physical state, $DK$, by 2 MeV, bringing it very close to the threshold, though it still does not form a bound state.

\begin{figure}[htbp]
		\centering
		\includegraphics[width=0.9\linewidth]{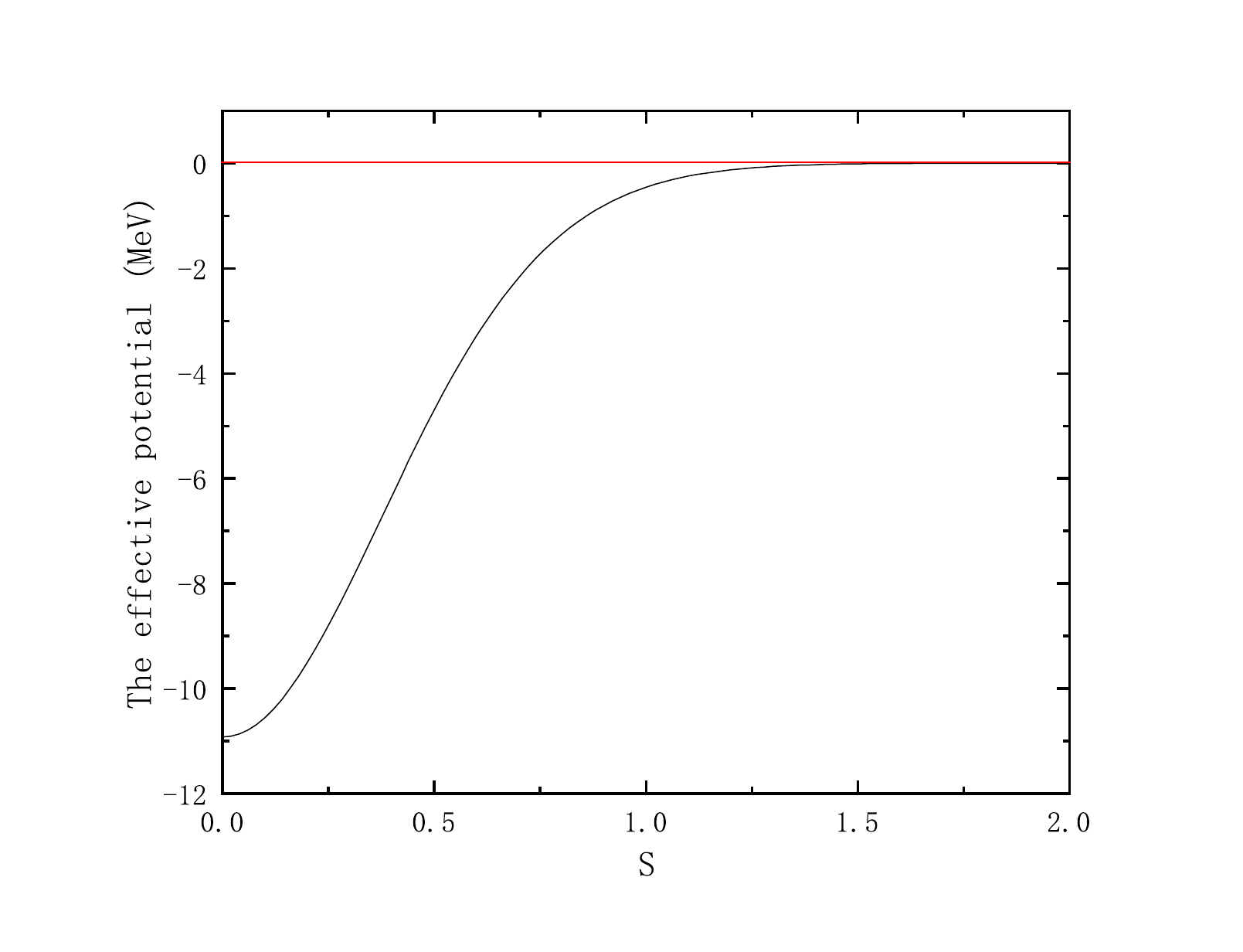}
		\caption{The effective potential of $DK$}
		\label{EF}%ÎÄÖÐÒýÓøÃͼƬ´úºÅ
\end{figure}

We observe that in our calculations, the $DK$ molecular state does not form a bound state, which contrasts with many previous studies. Therefore, in the following section, we explore the interaction between the $D$ and $K$ mesons by examining the equivalent potential between them. Specifically, the Gaussian wave function describing the relative motion between the $D$ and $K$ mesons is embedded into the distance vector $\vec{S}$, meaning that the original $\vec{r}$ becomes $\vec{r}-\vec{S}$. Consequently, Eq. \ref{radialpart} (where the wave function between $D$ and $K$ is an S-wave) becomes
\begin{equation}		
\label{radialpart2}		
\Psi(\mathbf{r,S}) = \left[\frac{2 \nu_n}{\pi}\right]^{\frac{3}{4}} e^{-\nu_{n}(r-S)^2}.		
\end{equation}
By varying the distance vector $\vec{S}$, we change the distance between the $D$ and $K$ mesons to obtain the energy variation curve for the $DK$ system and evaluate the equivalent potential between the $D$ and $K$ mesons. Here, we assume that the direction of $\vec{S}$ is the same as that of $\vec{r}$, so we only consider the magnitude of $\vec{S}$, denoted as $S$. According to our results, as $S$ increases, the energy of the system gradually increases until it approaches a constant. The results are shown in FIG. \ref{EF}, where the vertical axis represents the energy with the distance vector $S$ introduced, subtracted by the energy at infinite $S$ (in our calculation, the energy is considered to be effectively constant when $S > 2$). We observe that when $S$ is small, there is an attractive interaction between the mesons. However, no bound state is formed, possibly because at short distances (when $S$ is small), the repulsive kinetic energy is large enough to prevent bound state formation.

\subsection{The mixing of two quark and tetraquark}

%\begin{table}[t]
%\begin{center}
%\caption{Adjusted model parameters.\label{2317parameters}}
%\begin{tabular}{ccccc}
%  \hline \hline
%  ~$m_s$ (MeV)~         & ~$\alpha_{qs}$ ~         & ~$\alpha_{ss}$ ~       & ~$\alpha_{sc}$ ~  \\
%        ~~~~     441 ~~~~           &~~~~ 0.60 ~~~~          & ~~~~ 0.50   ~~~~       &~~~~ 0.49 ~~~~          \\
%  \hline \hline
%\end{tabular}
%\end{center}
%\end{table}

In this section, we mainly focus on the two-quark and four-quark mixing effects in $D_{s0}^{*}(2317)$, where we use the unquenched quark model to study the mass spectrum of the $D_s$ meson. In principle, the unquenched quark model requires re-adjusting parameters to incorporate the unquenched effects for all mesons, as these effects inevitably lower the meson masses, which would make the meson masses under the original model parameters inconsistent with experimental values. However, considering that the unquenched effects on the ground-state mesons, $D_s$ and $D_{s}^{*}$, are relatively small, we only adjusted the parameter $a_s$ in the spin-orbit interaction to ensure that the mass of $D_{s0}^{*}$ is above the $DK$ threshold, i.e., $a_s = 0.877$.

In the unquenched calculations for the ground-state mesons $D_s$ and $D_s^{*}$, we found that their overall mass shifts are relatively small, 23 MeV and 29 MeV, respectively, with the components mainly consisting of the bare $c\bar{s}$ quark pair. Since $D_s^{*}$ has quantum numbers of $1^{-}$, there are many possible four-quark states that contribute to this quantum number, which are listed in Table \ref{try}. Therefore, the mass shift of $D_s^{*}$ is relatively large, reaching 29 MeV.

For the $P$-wave mesons, the contribution from meson-meson channels plays an important role. The percentage of the $c\bar{s}$ component in $D^{*}_{s0}(2317)$ and $D_{s1}(2460)$ is $\sim 68\%$ and $\sim 67\%$, respectively. The main component of the $P$-wave meson is the $S$-wave $D^{(*)}K^{(*)}$, while other meson-meson channels have rather small contributions to the $D_s$ meson properties. Our results agree qualitatively with those of other works~\cite{31,33,34,35,36}. In the following, a detailed analysis of the two $P$-wave states is presented.

%
%Because the introducing of four-quark components to quark-antiquark states, the model parameters have to
%be adjusted, the adjusted parameters are listed in Table \ref{2317parameters}. The calculated results are shown in Table \ref{try}.
%From the table, one can see that the $S$- and $P$-wave $D_s$ mesons can be described well in the unquenched quark model.
%The dominant component for the ground states $D_s$ and $D_s^*$ is $c\bar{s}$, $\sim$ 97\%. which
%supports the validity of the quenched quark model in describing the ground state of hadrons.
%For the $P$-wave mesons, the contribution from meson-meson channels play an important role, the percentage of $c\bar{s}$ component in
%$D^{*}_{s0}(2317)$ and $D_{s1}(2460)$ is $\sim 46\%$ and $\sim 41\%$, respectively. The main component of $P$-wave meson is $S$-wave $D^{(*)}K^{(*)}$,
%other meson-meson channels have rather small contributions to $D_s$ meson properties.
%Our results agree qualitatively with that of other work~\cite{31,33,34,35,36}. In the following, a detailed analyses for two $P$-wave states
%are presented.
\begin{table*}[tp]
  \centering
  \fontsize{9}{8}\selectfont
  \makebox[\textwidth][c]{
   \begin{threeparttable}
   \caption{The masses of $D_s$ mesons and the percentages of $c\bar{s}$ component in the states. `S', `P' and `D' in the parentheses denote
   $S$-, $P$- and $D$-wave states.}\label{try}
    \begin{tabular}{ccccccccc}
    \hline    \hline
    \multirow{2}{*}{~} &\multicolumn{2}{c}{$D_s$}&\multicolumn{2}{c}{$D_s^*$}&\multicolumn{2}{c}{$D^{*}_{s0}(2317)$}&\multicolumn{2}{c}{$D_{s1}(2460)$} \\
    bare mass&\multicolumn{2}{c}{1953.6}&\multicolumn{2}{c}{2079.9}&\multicolumn{2}{c}{2361.8}&\multicolumn{2}{c}{2478.2} \\
    %\cmidrule(lr){2-3} \cmidrule(lr){4-5} \cmidrule(lr){6-7}
    ~& ~~~~$\Delta_{M}$~~~~& ~~~Per~~~&~~~$\Delta_{M}~~~$& ~~~Per~~~& ~~~$\Delta_{M}~$~~~&~~~Per~~~&~~~$\Delta_{M}$~~~&~~~Per~~~\\ \hline
    \midrule
     $DK^{*}(P)+D^{*}K(P)+D^{*}K^{*}(P)$             & $-20.1$& 1.3\% &  $-8.8$& 0.8\% & ~~~~-&~~~~~- & ~~~~-&~~~~~- \\
     $D_s\phi(P)+D_s^{*}\eta(P)+D_s^{*}\phi(P)$      & $-3.4$& 0.2\% &  $-1.4$& 0.1\% & ~~~~-&~~~~~- & ~~~~-&~~~~~- \\
     $DK(P)+D^{*}K^{*}(P)$                           & ~~~~-&~~~~~- &  $-3.1$& 0.4\% & ~~~~-&~~~~~- & ~~~~-&~~~~~- \\
     $D_s\eta(P)+D_s^{*}\phi(P)$                     & ~~~~-&~~~~~- &  $-0.5$& 0.1\% & ~~~~-&~~~~~- & ~~~~-&~~~~~- \\
     $D^{*}K^{*}(P)$                                 & ~~~~-&~~~~~- & $-13.4$& 1.0\% & ~~~~-&~~~~~- & ~~~~-&~~~~~- \\
     $D_s^{*}\phi(P)$                                & ~~~~-&~~~~~- &  $-2.5$& 0.2\% & ~~~~-&~~~~~- & ~~~~-&~~~~~- \\
     $DK(S)+D^{*}K^{*}(S)$                           & ~~~~-&~~~~~- & ~~~~-&~~~~~- & $-17.2$&30.1\% & ~~~~-&~~~~~- \\
     $D_s\eta(S)+D_s^{*}\phi(S)$                     & ~~~~-&~~~~~- & ~~~~-&~~~~~- &  $-0.6$& 0.1\% & ~~~~-&~~~~~- \\
     $D^{*}K^{*}(D)$                                 & ~~~~-&~~~~~- & ~~~~-&~~~~~- & $-17.9$& 1.5\% &  $-0.1$& 0.1\% \\
     $D_s^{*}\phi(D)$                                & ~~~~-&~~~~~- & ~~~~-&~~~~~- &  $-2.4$& 0.2\% &  $-0.1$& 0.1\% \\
     $DK^{*}(S)+D^{*}K(S)+D^{*}K^{*}(S)$             & ~~~~-&~~~~~- & ~~~~-&~~~~~- & ~~~~-&~~~~~  & $-22.0$&27.5\% \\
     $D_s\phi(S)+D_s^{*}\eta(S)+D_s^{*}\phi(S)$      & ~~~~-&~~~~~- & ~~~~-&~~~~~- & ~~~~-&~~~~~  &  $-0.8$& 0.1\% \\
     $DK^{*}(D)+D^{*}K(D)+D^{*}K^{*}(D)$             & ~~~~-&~~~~~- & ~~~~-&~~~~~- & ~~~~-&~~~~~  & $-18.3$& 4.5\% \\
     $D_s\phi(D)+D_s^{*}\eta(D)+D_s^{*}\phi(D)$      & ~~~~-&~~~~~- & ~~~~-&~~~~~- & ~~~~-&~~~~~  &  $-3.2$& 0.3\% \\
     Total                                           & $-23.5$&  1.5\%& $-29.7$&  2.6\%& $-38.1$& 31.9\%& $-44.3$&32.6\%\\ \hline
    Unquenched mass                                  &1930.1& 98.5\%&2050.2& 97.4\%&2323.7.8& 68.1\%&2433.9&67.4\%\\ \hline \hline
    \end{tabular}
   \end{threeparttable}}
  \end{table*}

\subsubsection{$D^{*}_{s0}(2317)$}

The $D^*_{s0}$ in quenched quark model is $c\bar{s}$ state with $IJ^{P}=00^{+}$, and the spin of the $c\bar{s}$ system must be 1, the spin-orbit
interaction plays a role. From the energy difference between $D^*_{s0}$ and $D_s^*$, $2317-2112 = 205$ MeV and the $P$-wave excitation energy
from kinetic energy, one can estimate the spin-orbit splitting in the $c\bar{s}$ state is around $205-492=-287$ MeV.
In unquenched quark model, the components of meson-meson states are $c\bar{q}q\bar{s}$ or $c\bar{s}s\bar{s}$ will enter the state $D^*_{s0}$.
The physical contents of $c\bar{q}q\bar{s}$ system can be $DK$ and $D^*K^*$, while $c\bar{s}s\bar{s}$
system can be $D_s\eta$ and $D_s^*\phi$.
In the Table \ref{try}, one can see that the $S$-wave $DK$, $S$- and $D$-wave $D^*K^*$ components occupies about $32\%$ of $D^*_{s0}(2317)$ contents,
which is in  qualitatively agreement with that of the other work \cite{28,31,33,34,35,36}. The contribution from
the other channels, such as $D_s^*\phi$, is very small due to the OZI rule violation. It is worth to mention that the $D$-wave $D^*K^*$ make a rather
large mass shift to the $c\bar{s}$ state although its percentage is not large. By using the contents of $D^*_{s0}(2317)$, we can estimate
the width of $D_{s0}^*(2317)\rightarrow D_s\pi$, 85 keV by averaging the width of bare $c\bar{s}$ (5 keV $\times$ 68\%) and molecular state $DK$
(130 keV $\times$ 32\%)~\cite{01,03,06,09,12,16}. The decay widths of $D^{*}_{s0}(2317)$ in bare $c\bar{s}$ and molecular $DK$ pictures are shown
in Table \ref{width}.

\begin{table}[t]
\begin{center}
 \caption{Decay width of $D^{*}_{s0}(2317)$ in bare $c\bar{s}$ and molecular $DK$ pictures. (unit: keV).\label{width}}
 \begin{tabular}{ccccc}
    \hline \hline \noalign{\smallskip}
                    & Ref. \cite{01} & Ref. \cite{03} & Ref. \cite{06} & Average\\
   bare $c\bar{s}$  &           10&          1.9&      2.4-4.7& 5\\    \hline
                    & Ref. \cite{09} & Ref. \cite{12} & Ref. \cite{16} & Average\\
   molecular $DK$   &   104-116&          180&      120     & 130\\
   \hline \hline
  \end{tabular}
 \end{center}
\end{table}

Although P.G. Ortega has used $^3P_0$ model to analysed $D^{*}_{s0}(2317)$ and  $D_{s1}(2460)$, we have some improvement in this work. Firstly,
the work~\cite{35} used the screening confinement in the unquenched model, this may lead to double-counting, because the screened confinement
has absorbed sea quark effect~\cite{BQLi}. Secondly, our modified $^3P_0$ model has considered two kind of corrected factor: one is distance
convergence and the other is energy suppression. Finally, all of the possible meson-meson channels are considered.
\subsubsection{$D_{s1}(2460)$}
The $D_{s1}$ in quenched quark model is $c\bar{s}$ state with $IJ^{P}=01^{+}$, because its energy is higher than $D_{s0}^*$, the spin of the state
is expected to be 0, there is no contribution from spin-orbit interaction. So we can estimate the $P$-wave excitation energy from kinetic energy,
$2460-1968=492$ MeV.
As shown in the Table \ref{mesonmass}, the mass of $D_{s1}^{\prime}$ in quenched quark model is close to the experimental value
of $D_{s1}(2460)$, and the quark content is $c\bar{s}$ with quantum numbers $IJ^P=01^+$, the unquenched masses is almost the same as that of quenched mass.
In unquenched quark model, the percentage of $c\bar{s}$ of $D_{s1}(2460)$ is about 67\%, and the main components of meson-meson channels
are $S$-wave $D^{(*)}K^{(*)}$. The contribution from $D_s\phi$, $D_s^*\phi$ and $D_s^*\eta^{\prime}$ is very small due to the OZI rule violation.

%\begin{table}[t]
%   \begin{center}
%   \caption{Values of $M(D^{*}_{s0}(2317))-M_{\bar{1S}}$, predicted by our quark model and the other work. The $M_{\bar{1S}}=\frac{1}{4}(D_s+3D_s^*)$ is the spin-averaged groundstate mass.\label{2317energy}}
%    \begin{tabular}{cccccc}
%    \hline\noalign{\smallskip}
%                 & Our work & Re\cite{35} & Re\cite{36}-1 & Re\cite{36}-2 & Exp.\\
%       bare mass & 287.8    & 309.0& 274.7  & 320.4  & 241.7\\
%unquenching mass & 258.6    & 249.6& 254.4  & 245.0  & 241.7\\\hline
%    \end{tabular}
%   \end{center}
%  \end{table}
%
%\begin{table}[t]
%   \begin{center}
%   \caption{Values of $M(D_{s0}(2460))-M_{\bar{1S}}$, predicted by our quark model and the other work. The $M_{\bar{1S}}=\frac{1}{4}(D_s+3D_s^*)$ is the spin-averaged groundstate mass.\label{2460energy}}
%    \begin{tabular}{cccccc}
%    \hline\noalign{\smallskip}
%                 & Our work & Re\cite{35} & Re\cite{36}-1 & Re\cite{36}-2 & Exp.\\
%       bare mass & 384.7    & 495.6& 383.3  & 398.0  & 383.3\\
%unquenching mass & 365.7    & 409.9& 377.4  & 392.0  & 383.3\\\hline
%    \end{tabular}
%   \end{center}
%  \end{table}

\section{Summary}

We investigate the exotic states $D^{*}_{s0}(2317)$ and $D_{s1}(2460)$ in the unquenched quark model with modified transition operator,
which works well in light meson spectrum \cite{qm5} and charmonium spectrum \cite{X3872_1}. All of calculation are done with help of Gaussian
expansion method.

Although our quenched quark model can describe $D^{*}_{s0}(2317)$ and $D_{s1}(2460)$ well in mass, the bare $c\bar{s}$ picture is difficult to
explain the large iso-violating decay, $D^{*}_{s0}(2317)\rightarrow D_s \pi$. The molecular $DK$ structure is invoked to explain the
structure of $D^{*}_{s0}(2317)$. So it is natural to study $D^{*}_{s0}(2317)$ in unquenched quark model. Our calculation shows that the main component
of $D^*_{s0}(2317)$ is $DK$ (32\%), so $D^*_{s0}(2317)$ is a complete mixture structure. On the other hand, with the composition of the state,
we roughly estimate the iso-violating decat width of $D^{*}_{s0}(2317)$ about 85 keV. For $D_{s1}(2460)$, we have similar results.

In conclusion, $D^{*}_{s0}(2317)$ and $D_{s1}(2460)$ may be complete mixture structure of $c\bar{s}$ and molecular states($DK$ and $D^*K$)
in our modified unquenched quark model, while $S$-wave partners, $D_s$ and $D_s^{*}$ may be almost pure $c\bar{s}$ structure.

\acknowledgments{ This work is supported partly by the National Natural Science Foundation of China under Grant Nos. 12205249, 12205125, 11865019, 11775118 and 11535005 by the Natural Science Foundation of Jiangsu Province (BK20221166), the Science and Technology Foundation of Bijie (Grant No.BiKeLianHe[2023]17), and the Funding for School-Level Research Projects of Yancheng Institute of Technology (No. xjr2022039).}

\end{document}